\documentclass[12pt]{article}
\usepackage{amssymb,bbold}

\newcommand{\be}[3]{\begin{equation}  \label{#1#2#3}}
\newcommand{\ee}{\end{equation}}
\newcommand{\ba}{\begin{array}}
\newcommand{\ea}{\end{array}}
\newcommand{\bea}[3]{\begin{eqnarray}  \label{#1#2#3}}
\newcommand{\eea}{\end{eqnarray}}

\let\LARGE=\Large
\let\Large=\large
\let\large=\normalsize

\def\1{\mathbb 1}

\def\pa{\partial}

\begin{document}

\thispagestyle{empty}

\begin{flushright}
\hfill{AEI-2003-079}\\
\hfill{hep-th/0310090}

\end{flushright}

\vspace{15pt}

\begin{center}{ \LARGE{\bf
Chronological structure of a G\"odel type universe\\[4mm] 
with negative cosmological constant 
}}

\vspace{50pt}

{\bf Klaus Behrndt} \quad and \quad
{\bf Markus P\"ossel}

\vspace{20pt}

{\it  Max-Planck-Institut f\"ur Gravitationsphysik,
Albert Einstein Institut\\
Am M\"uhlenberg 1,  14476 Golm,
 Germany}\\[1mm]
{E-mail: behrndt@aei.mpg.de \quad , \quad 
mpoessel@aei.mpg.de}
\vspace{10pt}

\vspace{50pt}

{ABSTRACT}

\end{center}

\noindent
We show that the G\"odel solution of five-dimensional gauged supergravity
contains either closed time-like curves through every space-time point
or none at all, dependent on the rotational parameter.  In addition,
we present a deformation of that solution with a
parameter $\kappa$ which characterizes the symmetry of four-dimensional
base space: for $\kappa = 1 , 0, -1$ it has spherical, flat and hyperbolic
symmetry, respectively.  Also investigated are the causal properties of the
lifted solution in 10 dimensions.

\vfill

\newpage


\noindent
The occurence of closed time-like curves (CTCs) is 
widely viewed as a pathological
feature of space-time.  While there certainly are solutions of general
relativity that contain CTCs (a review of the classical examples can be
found in \cite{310}), there is an on-going discussion whether or not
a more complete description, in particular one that includes quantum
effects, would lead to the absence of CTCs \cite{330,240,320}.  
One widely-held view is that a viable theory of quantum gravity should not 
permit their existence, at least not as far as curves
accessible to an outside observer are concerned (as opposed to, say, the CTCs
hidden behind the horizon of a rotating black hole). However,
in recent time it was realized that string theory, as
one candidate for a theory of
quantum gravity, does allow for vacua with ``naked'' CTCs,
and in particular that some such vacua preserve at least one supersymmetry
and thus can be expected to be reasonably stable.

The first such example was the rotating black hole solution in 5
dimensions \cite{140,130} where, for sufficiently large angular
momentum, CTCs appear {\em outside} the horizon \cite{170,070,350}.
If one embeds such black hole into an asymptotic Anti-de Sitter space
\cite{040}, it acquires an ergosphere which contains CTCs for
arbitrary values of the angular momentum \cite{180}. Another class of
string solutions are supertubes, where the space-time contains CTCs in
the over-rotating case \cite{050}, analogous to the asymptotically
flat black holes.  There is an on-going discussion about whether one
should exclude these vacua in string theory, or at least show that the
pathological regions do no harm -- either since they somehow cannot be
reached by geodesic motion or because a more fully string-theoretical
description indicates the need to excise and to replace them with
non-pathological space-time patches.  More specifically, a typical
test involves the examination of these backgrounds using different
kinds of brane probes and investigates these probes' world volume
theory.  Indications that the probes cannot reach the pathological
regions (behind the ``CTC horizon'') include the appearance of ghost
degrees of freedom (with negative kinetic energies) in the
world-volume theory, see for example \cite{070,260}.  Another approach
has examined the question of whether a problematic black-hole
space-time containing CTCs can be created in some way from CTC-free
space-times, using the string-theoretical microscopic description for
the black hole in question \cite{090}. However, while certainly
valuable, arguments of this type are limited insofar as they do not
address ``eternally existing'' pathological configurations.  (Another
possible problem with the argument in the reference cited is its
reliance on an energy condition that might be violated in quantum
gravity).

Yet another class of chronology violating solutions, and one which has
attracted much attention recently, is the G\"odel-type solution of
5-dimensional ungauged supergravity, where the matter of the original
G\"odel solution is replaced by a magnetic configuration of the
Abelian gauge field \cite{030}. This homogeneous solution describes a
rotating universe and has CTCs through every point of space-time.  A
number of interesting properties of this solution has been explored:
The CTCs can be shielded by holographic screens, and the solution is
T-dual to certain pp-waves \cite{080,340}; black holes can be embedded
into this space-time \cite{070,060,250,370,220}, and it has been
examined using various probes \cite{110,100}.  Another interesting
aspect dates back to the original discovery of the solution, which can
be obtained by dimensional reduction of a WZW model discussed by Nappi
and Witten \cite{200} that is itself a special case of a larger class
of conformal field theories introduced in \cite{190}.  Interestingly
enough, these CFTs can be solved exactly \cite{210}.  The fact that
the partition function of this background becomes divergent has been
linked to the appearance of CTCs and interpreted as strings becoming
tensionless \cite{270}, however, a physical understanding of this
phase transition is still missing.  Let us also mention that there is
an interpolating string solution between this solution and the exact
flux backgrounds of the Melvin type \cite{230}, which exhibits similar
instabilities \cite{280}.

In \cite{020} the corresponding G\"odel solution of minimal gauged
supergravity was presented.\footnote{
The four-dimensional cousin of this
space-time was recently found in \cite{300}.  
} This solution also exhibits 
CTCs, but these have not so far been
discusssed in detail.  It solves the Einstein-Maxwell equations 
associated with
the bosonic Lagrangian of minimal five-dimensional supergravity,
\be010
*{\cal L} \sim *1 \Big[ {R\over 2} + \chi^2 - {1 \over 4} F^2\Big]
 - {2 \over 3 \sqrt{6}}\:  F \wedge F \wedge A \ ,
\ee
in which
the cosmological constant $\chi^2$ appears due to a
Abelian gauging of the R-symmetry $SU(2)$ .  By construction within
the framework of \cite{020}, the solution is supersymmetric.  It belongs
to a class of solutions 
defined by a stationary metric with a K\"ahler base space;
preservation of supersymmetry entails that the field strength of the 
Abelian gauge field be anti-selfdual over the K\"ahler space, and have no
components in the time-direction.  In the form discussed in 
\cite{020}, the G\"odel-type solution is based on the Bergmann metric 
parameterizing the coset manifold ${SU(2,1) \over U(2)}$, its
five-dimensional metric and gauge field are\footnote{Compared with 
the form given in \cite{020}, we have rescaled field strength and 
cosmological constant, chosen the (irrelevant) phase angle between 
${\cal F}_1$ and ${\cal F}_2$ to be zero, and performed a coordinate 
redefinition $\varrho=6/\chi^2\sinh^2(\chi r/\sqrt{6})$ to remove 
cumbersome hyperbolic functions.  }
\be100
ds^2 = - ( dt + \omega)^2 + {d \varrho^2 \over 4 \varrho V(\varrho)}
+ {\varrho \over 4} \, \Big[ V(\varrho) 
\sigma_3^2 + \sigma_1^2 + \sigma_2^2 \Big] 
\mbox{\quad and \quad}
A = {\sqrt{3} f \varrho \over 8 V(\varrho)} \, \sigma_1 
\ee
where
\be101
\omega = {\chi \, \varrho \over 2 \sqrt{6}} \, \sigma_3 + {f \, \varrho \over
4 \sqrt{2} V(\varrho)}\,  \sigma_1 \quad , \qquad 
V(\varrho) := 1 + {\chi^2 \over 6} \varrho
\ee
and where the left-invariant $SU(2)$ 1-forms
\be020
\ba{l}
\sigma_1 = \sin\phi \, d\theta - \cos\phi \sin\theta \, d\psi\ , \\
\sigma_2 = \cos\phi d\theta +\sin \phi \sin\theta d\psi \ , \\
\sigma_3 = d \phi + \cos\theta \, d\psi \
\ea
\ee
were used in order to specify our solution in terms of Eulerian coordinates
$0\leq\rho, 0\leq\theta\leq \pi, 0\leq \psi\leq 2\pi$ and
$0\leq \phi< 4\pi$.
In the limit of vanishing cosmological constant, $\chi\to 0$, one
recovers the supersymmetric G\"odel solution, which, in Cartesian coordinates,
reads \cite{030}
\be552
\ba{rcl}
ds^2 &=& - ( dt +  \omega )^2 + dx^m dx^m \\
\omega = {2 \over \sqrt{3}} A &=&  2 \gamma\, (
x^1 dx^2 - x^2 dx^1 + x^3 dx^4 - x^4 dx^3) \ .
\ea
\ee
Regarding the $\chi=0$ solution, the proof that 
there are CTCs through every point
is greatly helped by the fact that the space-time
is homogeneous. Once
the existence of any CTC, however special, has been proven, homogeneity
ensures that CTCs exist everywhere.  This argument does not carry over
to the G\"odel solution in gauged supergravity, however, where homogeneity
is lost.  So far, it has not yet been shown whether there
are regions of this space-time that are free of CTCs. As we will discuss in
detail now, this is unfortunately not the case.

Before this discussion, however, let us present a brief collection of further
information about the gauged five-dimensional G\"odel universe.  First of all,
there is an easy generalization of the solution in the following way:
For the base space $SU(2,1)\over U(2)$, different parametrizations
are possible.  In the case of the Bergmann metric, the base metric
is a complex line bundle over $S_2$; as a generalization, one can admit
a complex line bundle over other two-dimensional spaces of constant
curvature $\kappa$.  This corresponds to the $n\rightarrow \infty$ limit 
of the metric described in \cite{290} and from this base-space
five-dimensional solutions are readily constructed, namely
\be362
\ba{rcl}
ds^2 &=& -(dt + \omega)^2 + {d\varrho^2 \over 4 \varrho V(\varrho)}
+ {\varrho \over 4}  V(\varrho) \Big[ d\phi' + {1 \over 2} {xdy -ydx \over
    1 + {\kappa \over 4} (x^2 + y^2)}\Big]^2 + {\varrho \over 4}
    {dx^2 + dy^2 \over [1 + {\kappa \over 4} (x^2 + y^2)]^2} \\[2mm]
A &=& {\sqrt 3 f \over 8}\, \Big[ {\varrho  \over \kappa V(\varrho) 
[1 + {\kappa \over 4} (x^2+y^2)]}  ( \sin\kappa\phi' \, dx
+  \cos \kappa\phi' \, dy )  
- {6 \over \kappa \, \chi^2}  dy \Big],
\ea
\ee
where $\kappa$ can be scaled so that three different cases 
$\kappa = -1, 0, 1$ remain.  In this definition,
\be152
\omega = {\chi \, \varrho \over 2 \sqrt{6}} \,
\Big[ d\phi' + {1 \over 2} {xdy -ydx \over
    1 + {\kappa \over 4} (x^2 + y^2)}\Big]  + A
 \quad , \qquad
V(\varrho) = \kappa + {\chi^2 \over 6} \varrho \ .
\ee
(Note that we have added a constant part to $A_y$ to ensure a smooth limit
$\kappa=0$.)  In order to avoid a conical singularity, one should require
a periodicity $\phi' \simeq \phi' + {4 \pi \over \kappa}$; in particular,
for $\kappa = 0$ all coordinates are non-compact.  Let us note
here another possible choice for the gauge connection, namely
\[
A_x + i A_y = - {1 \over \kappa} \log \Big[ \Big( {6 \kappa  \over 
\chi^2\varrho} + 1\Big)
 \Big(1 + {\kappa \over 4} [x^2 + y^2]\Big) e^{i\phi'}\Big]
(dx + i dy) \ .
\]
The ambiguity in the choice of gauge connection can be understood
as follows:  The (complex) field strength two-form is 
conformally equivalent to the two non-exact two-forms of the 
quaternionic space ${SU(2,1) \over U(2)}$, with the
conformal factor defined by the K\"ahler potential.  However,
the split into an exact two-form and a pre-factor is not unique, and
it might be interesting to explore the possible relation of this
to K\"ahler transformations.

Having noted the existence of this one-parameter family of solutions,
we now focus on the spherical case  $\kappa = 1$.  It corresponds to the
G\"odel-type solution discussed before, and the 
metric (\ref{100}) can be recovered changing coordinates
as $ x + iy = 2 \, e^{i\psi} \, \tan {\theta/2}$.  

For the discussion of CTCs, the best starting point are left-and 
right-invariant vectors associated with the Euler angle coordinates --
natural candidates for tangent fields of congruences of closed curves.
In flat space with standard metric, these are all Killing vectors;
constructing the space-time with arbitrary values of $\chi$ and $f$,
however, most of them loose their Killing property (of course, other
Killing vectors are gained, notably $\xi_0=\partial_t$ associated with
the stationarity of the space-time).  First of all, there are the
$SU(2)$-right-invariant vectors
\be120
\ba{l}
\xi_1^R = -\sin\psi\:\partial_{\theta} 
          -\cot\theta\:\cos\psi\:\partial_{\psi}
          +\frac{\textstyle\cos\psi}{\textstyle\sin\theta}\:\partial_{\phi}\\
\xi_2^R = \phantom{-}\cos\psi\:\partial_{\theta} 
          -\cot\theta\:\sin\psi\:\partial_{\psi}
          +\frac{\textstyle\sin\psi}{\textstyle\sin\theta}\:\partial_{\phi}\\
\xi_3^R = \partial_{\psi}.
\ea
\ee
These all remain Killing vectors of the full space-time.  Secondly,
there are the $SU(2)$-left-invariant vectors
\be130
\ba{l}
\xi_1^L = -\frac{\textstyle\cos\phi}{\textstyle\sin\theta}\partial_{\psi}
+\sin\phi\:\partial_{\theta} + \cot\theta\:\cos\phi\:\partial_{\phi}\\
\xi_2^L = \cos\phi\:\partial_{\theta}
         +\frac{\textstyle\sin\phi}{\textstyle\sin\theta}\:\partial_{\psi}
         -\cot\theta\:\sin\phi\:\partial_{\phi}\\
\xi_3^L = \pa_{\phi}.  
\ea
\ee
Dual as they are to the left-invariant forms used in the construction of
the metric, the products between these vectors are comparatively 
simple, and we will thus use these vectors below in the construction of
CTCs.  They are, however, Killing only in special limits:  $\xi_3^L$
becomes a Killing vector in the limit $f=0$, for arbitrary $\chi$, when
the gauge field is switched off, the Einstein equations reduce to
$Ric=-\frac23\chi^2 g$ and the space-time becomes Anti-deSitter
in unusual coordinates (to return to static coordinates, one need merely
shift $\phi \rightarrow \phi'=\phi- {2 \chi \over \sqrt{6}} t$).
On the other hand, $\xi_1^L$ becomes a Killing vector in the 
limit $\chi=0$ of the non-gauged G\"odel solution.

With these preparations, now for an analysis of the possible CTCs
in the space-time (\ref{100}).  As it turns out, the presence or absence of
CTCs in that space-time is governed by the relative values of the
parameters $\chi$ and $f$: As long as 
\be652
f^2\; \leq \;  {4 \over 3} \; \chi^2
\ee
there are no CTCs; otherwise, there are CTCs through every space-time point.  
(Note that this is a more stringent limit than the one that resulted
from the construction of special CTCs in \cite{020}; the limit given
there would, in our 
conventions\footnote{Notably 
${\cal F}_1^2+{\cal F}_2^2 =f^2/2$ and $\chi_{there}=
\sqrt{2}\cdot\chi_{here}$.}
be $f^2\leq 16/3\chi^2$.)  

This can be seen as
follows.  First, we can ask ourselves under what conditions the
coordinates chosen in (\ref{100}) define a global foliation of the
space-time they describe, in other words: under what conditions is the
hypersurface $\Sigma_t$, defined by $\rho,\theta,\psi,\phi$ at
$t=const.$, spacelike everywhere?  The best way to find out is a look at the
eigenvalues of the induced metric on each such hypersurface.  Doing so
(a basis of $\pa_t,\pa_{\rho},\xi_1^L,\xi_2^L,\xi_3^L$ proves
convenient), we see that it has three eigenvalues that are positive at
any $\rho$ for arbitrary values of $f$ and $\chi$ (zeroes at $\rho=0$
turn out to be coordinate artefacts, as can be seen by changing to
cartesian coordinates).  Not so for the fourth eigenvalue: As long as
$f^2>4/3\chi^2$, it starts out as positive for small $\rho$, has a
zero at $\rho_0=24/(3f^2-4\chi^2)$ and is negative
for all larger values of $\rho$.  This tells us directly that, as long
as $f^2\; \leq \;  {4 \over 3} \; \chi^2$
all is well, and the space-time contains no CTCs. Viewing
this as a bound on the parameter $f^2$, which measures the angular
momentum/the magnetic flux, the situation is somewhat similar
to the rotating black holes, with the presence of CTCs governed by 
angular momentum versus total mass. 

Next for the proof that, as long as $f^2>4/3\chi^2$, there are CTCs
through {\em every} space-time point.  First, let us look at CTCs that
remain at constant coordinate distance $\rho=const.$ from the origin.  In order
to find such curves, it is no longer sufficient to consider only
integral curves along one suitably chosen angular coordinate, as was
possible in the non-gauged case \cite{030} or in the case of
supersymmetric, five-dimensional black holes \cite{180}.  Instead, a
more general Ansatz is needed.  Our above analysis of the induced
metric on the hypersurfaces $\Sigma_t$ suggests a likely candidate for
a tangent vector field build from the $\xi_i^L$, namely the
eigenvector associated with the eigenvalue that is responsible for the
degeneracy of the metric of $\Sigma_t$ at $\rho=\rho_0$,
\be210 
{\xi}_c = \left[\sqrt{1+\frac{3f^2}{16\chi^2V(\rho)^2}}-\frac{\sqrt{3}f}{4\chi
    V(\rho)}\right]\xi_3^L +\xi_1^L.
\ee
It is readily checked that $|\xi_c|^2<0$ for $\rho\geq\rho_0$, as long as
$f^2>4/3\chi^2$.
It still needs to be shown that the congruence defined by the tangent
field $\xi_c$ consists of {\em closed} curves.  This can be seen as follows:
As manifolds, the surfaces $\rho=const.$ are homeomorphic to
three-spheres.  If any such three-sphere is equipped with the standard
metric, then the $\xi^L_i$ are Killing vectors, and the geodesics are
great circles.  In particular, the projection of the tangent vector
onto any $\xi^L_i$ is constant along each geodesic so, as the
$\xi^L_i$ form a basis field for the tangent space of the $S^3$, each
geodesic is an integral curve of some linear combination of the
$\xi^L_i$ and, conversely, the integral curve of any such linear
combination is closed.

So far, we have shown that, for suitable $f$, CTCs exist in some regions
of space, namely for $\rho>\rho_0$.  The fact that CTCs pass through all 
points of the region $\rho\leq\rho_0$, as well is plausible, as
those closed timelike curves whose existence we have shown should be easily
deformable to ``timelike winding stairs'', i.e.\ open curves that lead from
a space-time point $(t,\rho,\theta,\psi,\phi)$ to
$(t-\Delta t,\rho,\theta,\psi,\phi)$ for some finite $\Delta t$.  Given such
spiral curves, we can travel from $\rho<\rho_0$ into the region
$\rho>\rho_0$ in an ordinary time-like curve, descend the spiral to some 
suitably earlier coordinate time, and return to our initial $\rho$ value and
our initial coordinate time.  More concretely, taking as one of a plethora 
of possibilities the Ansatz\footnote{This is the above eigenvector 
$\epsilon_c$, with the coefficient of $\xi_3^L$ expanded around 
$\rho_0$ and evaluated at an arbitrarily chosen point 
$\rho=\rho_0+2/\chi^2$.  It has the advantage of being simple 
enough for $|\xi_{ws}|^2<0$ to be demonstrable with modest effort.}
\be220
\xi_{ws}= \frac{2\chi}{\sqrt{3}f}\left[1
       +\frac{(3f^2-4\chi^2)^2}{9f^2(3f^2+4\chi^2)}\right]\xi_3^L+\xi_1^L
       -c\:\pa_t.
\ee
With this definition, $|\xi_{ws}|^2$ is a quadratic function of the as
yet unfixed coefficient $c$ with two zeroes $c_{\pm}$.  Both for
$c\ll 0$ and for $c\gg 0$ we find $|\xi_{ws}|^2<0$ as expected:
choosing $c\ll 0$, we will end up near-parallel to $-\pa_t$, and
thus in the past light-cone; choosing $c\gg 0$, near-parallel to $\pa_t$
and in the future light-cone.   The necessary and
sufficient condition for $\xi_{ws}$ to be tangent to a winding stair
curve is for the smaller zero, $c_-$, to be {\em positive}.
Calculation of $|\xi_{ws}|^2$ at, for instance, $\rho=2\rho_0$ shows
this to be the case as long as $f^2>4/3\chi^2$, and choosing
$c=c_-/2>0$ defines an example of a vector field tangent to a
congruence of winding-stair curves.
This completes the proof of our statement that every point in
space-time is on a CTC as long the cosmological constant is small, but
in contrast to the case of vanishing cosmological constant all CTC
disappear in the moment the bound (\ref{652}) is saturated.

Finally, it is of interest to lift the five-dimensional to
string-friendly ten dimensions; in the case of black holes or the
G\"odel solution without cosmological constant, this lifting offers a
way to get rid of the offending CTCs \cite{360,070}.  More concretely,
we lift to a solution of ten-dimensional type IIB-supergravity
following the prescription given in section II of \cite{010}.  Let the
additional five (angular) coordinates be $\alpha,\beta$ and $\phi_i$,
$i=1,2,3$, and define functions $\mu_i$ of $\alpha,\beta$ such
that\footnote{For concreteness, one can choose $\mu_1 = \sin\alpha,
\mu_2 = \cos\alpha\,\sin\beta, \mu_3 = \cos\alpha\,\cos\beta$.}
$\sum_{i=1}^{3}\mu_i^2=1$.  The lifted metric constructed from
five-dimensional line element $ds_{5}^2$ and the one-form $A$ is
\be401
ds_{10}^2 = ds_{5}^2 +\frac{6}{\chi^2}
\sum_{i=1}^3\left[
(d\mu_i)^2 +\mu_i^2\left(d\phi_i+\frac{\chi}{3}A\right)^2
\right]
\ee
and, together with the five-form $\bar{F}=G+*G$ defined by
\be410
G = \frac{1}{2\sqrt{5}}\left[
\frac{\chi}{3}\varepsilon^{(5)} - \frac{1}{2\chi^2}\sum_{i=1}^3
d(\mu_i^2)\wedge d\phi_i\wedge*^{(5)}F
\right],
\ee
where $\varepsilon^{(5)}$ and $*^{(5)}$ are the five-dimensional 
volume form and Hodge dualization, respectively. It fulfills the
ten-dimensional Einstein-Maxwell-equations
\be500
Ric^{(10)}_{MN}-5\bar{F}_{MM2\cdots M_5}\bar{F}_N{}^{M2\cdots M_5}=0 \quad
\mbox{ and } \quad d*F = 0.
\ee
One might have hoped that this lifting alone could cure the illness of
the CTCs, as it involves a shift in the $U(1)$ fibre that is at the root
of the five-dimensional solution's chronological problems.  However,
this is not the case. Again, by looking at the foliation defined
by the given coordinates, it follows that for $f^2\leq 4/3\chi^2$, there are
no CTCs, 
and again, for $f^2> 4/3\chi^2$, CTCs through every space-time point can
be constructed as follows, in perfect analogy to the procedure we have
used in five dimensions:  To show the existence of CTCs through all
points in the region
$\rho > \rho^{(10)}_0=12/(\chi[\sqrt{3}f-2\chi])$, 
one can use the  tangent vector field $\xi_3^L+\xi_1^L$.  For the remaining
regions, study of the tangent field $-c\pa_t + \xi_3^L+\xi_1^L$ shows
for $f^2\leq 4/3\chi^2$ and some $\rho_{ws}>\rho_0$ the existence of
``winding stair'' curves that can be used to go back in time and hence
to construct CTCs of finite length through space-time points with
$\rho<\rho_0$.

There are a number of open questions which we were not able to
discuss in detail in this note, but which might be useful for future work.  
Notably, the G\"odel solution we examined can be seen as a deformation of 
$AdS_5$ and, as we have shown, CTCs exist even close to the boundary. 
This raises the question of what happens in the dual field theory. 
As we have mentioned, for the G\"odel solution in ungauged supergravity there
exist a description in string theory in terms of a solvable CFT; it is
natural to ask whether the solutions of gauged supergravity discussed
here can be described by an exact model.



\subsection*{Acknowledgments}

We would like to thank J\"urgen Ehlers and Arkady Tseytlin for useful
discussions.  The work of K.B.\ is supported by a Heisenberg grant of
the DFG.




\providecommand{\href}[2]{#2}\begingroup\raggedright\endgroup

\end{document}